\documentclass[lnicst]{svmultln}
\usepackage{makeidx}  % allows for indexgeneration
\usepackage[pdftex]{graphicx}
\DeclareGraphicsExtensions{.pdf,.jpeg,.png}
\usepackage[tight,footnotesize]{subfigure}
\usepackage{tabulary}
\usepackage{url}
\urldef{\mailsa}\path|{silvia.puglisi,monica.aguilar,jforne}@entel.upc.edu|
\urldef{\mailsb}\path|{gmarrugat,angel.torres.moreira}@gmail.com|

\begin{document}
\mainmatter              % start of the contribution
\title{MobilitApp: Analysing mobility data of citizens in the metropolitan area of Barcelona}
\titlerunning{MobilitApp}  % abbreviated title (for running head)
\author{Silvia Puglisi \and \'Angel Torres Moreira \and
 Gerard Marrugat Torregrosa \and M\'onica Aguilar Igartua \and Jordi Forn\'e}
\authorrunning{Silvia Puglisi et al.}   % abbreviated author list (for running head)
% \tocauthor{Ivar Ekeland, Roger Temam, Jeffrey Dean, David Grove,
% Craig Chambers, Kim B. Bruce, Elisa Bertino}
%
\institute{{Department of Network Engineering, \\
Universitat Polit\`ecnica de Catalunya (UPC)\\
C.\ Jordi Girona 1-3, 08034 Barcelona, Spain}\\
\mailsa\\
\mailsb\\
}

\maketitle              % typeset the title of the contribution
% \index{Ekeland, Ivar} % entries for the author index
% \index{Temam, Roger}  % of the whole volume
% \index{Dean, Jeffrey}

\begin{abstract}        % give a summary of your paper

MobilitApp is a platform designed to provide smart mobility services in urban areas. It is designed to help citizens and transport authorities alike. Citizens will be able to access the MobilitApp mobile application and decide their optimal transportation strategy by visualising their usual routes, their carbon footprint, receiving tips, analytics and general mobility information, such as traffic and incident alerts. Transport authorities and service providers will be able to access information about the mobility pattern of citizens to offer their best services, improve costs and planning.
The MobilitApp client runs on Android devices and records synchronously, while running in the background, periodic location updates from its users. The information obtained is processed and analysed to understand the mobility patterns of our users in the city of Barcelona, Spain.

% please supply keywords within your abstract
\end{abstract}

\section{Introduction}
Mobility and transportation efficiency have always been essential in a city for it to function properly. When the architect Ildefons Cerd\`a  i Sunyer (Spanish urban architect, 1815-1876) drafted his original plan of the extension of Barcelona in the 1850's, he focused on certain key points. Among these the need for seamless movement of people, goods, energy, and information. The extension of the city of Barcelona was conceived along the idea of ensuring more fluid traffic in all directions, above all for public transport. Cerd\`a wanted to make sure that the steam tram could circulate easily and used its long turning radius to determine the angle of the corners of the buildings.
\\
Smart mobility solutions can provide efficient, safe and comfortable transport services, so that visitors and residents can easily travel across the city. Smart mobility services are information driven and rely on technology to provide personalised services to its users. We present a smart mobility platform designed both for urban citizens and transport service providers alike. Our platform is composed by an analytics solution and a client application. The mobile application runs on Android devices, collecting mobility data regarding activities performed by its users. Smart citizens use the application to receive information regarding: their usual routes, their carbon footprint and traffic and incident alerts. Transport authorities can use the analytics solution to provide a clean, efficient and affordable transportation system, analyse mobility pattern of citizens, forecast and prevent network congestions, and general planning and execution.

\subsection{State of the art}
The flow of people and goods over the transport network of a city is a complex problem. Transport service providers need to forecast demand around the city and plan long term investment. To provide a more efficient transportation service, authorities in the field analyse data about the mobility patterns of users in metropolitan areas. These analysis are often conducted on partial ticket sales data, surveys and economical models. This information lacks reliability, since it only contains partial mobility samples, not representative enough of a whole metropolitan region.
Smart mobility programs and the use of smartphones offer a more efficient and precise way to collect data about transportation usage, while also providing citizens with comprehensive personalised services.
\\
Mobility patterns recognition is the problem of detecting the current mode of transportation of a person. Our approach is to use smartphones to detect mobility patterns. Studies implementing novel accelerometer-based techniques for accurate and fine-grained detection of transportation modes on smartphones~\cite{hemminki2013accelerometer} have shown promising results to capture key characteristics of vehicular movement patterns.  Unfortunately to recognise specific human activities and provide contextual information, a mobile application needs to continuously listen to sensor data from a user's mobile device. Several sensor readings need to be interpreted to produce meaningful data. Frameworks for human activity recognition (HAR) and context aware applications maximising power efficiency have been developed~\cite{sivakumar2015battery} leveraging on topics such as user profile adaptability and variant sensory sampling operations. Activity recognition algorithms and their power consumption, are also further improved by identifying spurious events classification and subsequently pruning a decision tree model on these specific cases~\cite{phan2014improving}. These approaches have shown a 10\% classification improvement in some occasions.
\\
With initiatives like open data, and rising development of open smart city services, urban areas have started to evolve into \emph{open source spaces} where it is possible to design new applications to integrate with existing technologies. Software development kits (SDKs) to develop mobile sensing apps and collect data about smartphones users are already part of most mobile operating systems~\cite{yang2014harlib}~\cite{funf}~\cite{detectedactivity}, turning smart citizens into actual sensors.
\\
It is important to create a mutual relationship between smart city services providers and the actual sensing citizens that contribute with their mobility patterns. More useful and better services are offered, so that the city can gather the necessary data to be more efficient~\cite{palazzi2010path}. The importance of smart cities in the end is not merely about efficiency and smart services, as much as enabling people to shape the urban environment. One aspect of this is providing platforms where people would be able to communicate and collaborate. An example of these applications are hyperlocal news and real times updates, connected urban mobility, and what is called grassroots design and city hacking~\cite{weber2014my}~\cite{townsend2013smart}.

\section{Platform architecture}
MobilitApp~\cite{mobilitapp}~\cite{mobilitapp-play} is a smart city platform able to obtain mobility data of the citizens in the metropolitan area of Barcelona, Spain. Our implementation synchronously collects updated geographical position as well as users' current activities. At the end of the day, information is processed and sent to our backend where it is stored and analysed.
Our platform, part of the INRISCO project, is designed specifically to provide smart tools to transport authorities and is developed with feedback from ATM (Autoritat del Transport Metropolit\`a). More specifically the MobilitApp platform provides the following functionalities:

\begin{itemize}
 \item Traffic information in real-time
 \item Traffic incidences
 \item Web app to analyse and filter the collected information on mobility.
\end{itemize}

\begin{figure}[t]
\centering
\includegraphics[width=90mm]{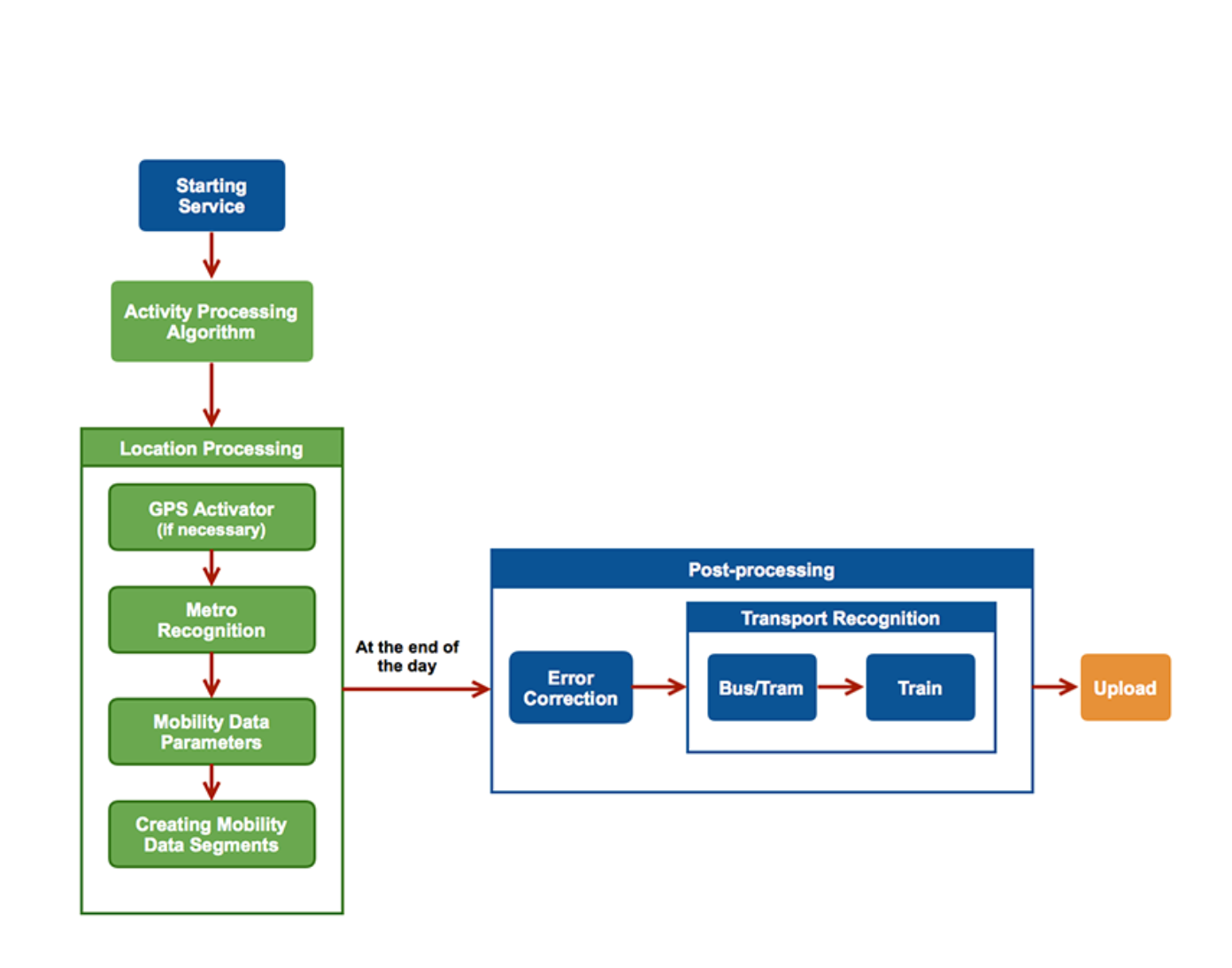}
\caption{How users' activities are collected and processed in our MobilitApp application.
\label{processing}}
\end{figure}

\subsection{Mobile application}
MobilitApp collects citizens' mobility patterns in the background. MobilitApp uses Google Android APIs to discover the user's positions. This can be considered as a first approach for user's activity detection.  The APIs provide a low consumption mechanism to log periodic updates and detected activity types by using mainly the device accelerometer. This information is then processed sample by sample to increase accuracy and efficiency. Our own implementation of the activity detection algorithm consists in calling the Activity APIs and sampling the obtained results every 20 seconds. Then every 2 minutes, the algorithm makes a statistical estimation of the most probable result out of the last samples.
\noindent
To further polish the results obtained we also consider the following factors:
\begin{itemize}
 \item Accuracy of the GPS: when a device is underground, the GPS accuracy decreases consistently.
 \item Location of points of interest(POI) to help the algorithm knowing if a user is \emph{close} to a bus stop or a metro station.
 \item Directions: we use Google Directions APIs to check if there is a known route (using all possible transportation types) between two points.
\end{itemize}

\begin{figure}[t]
\centering
\includegraphics[width=100mm]{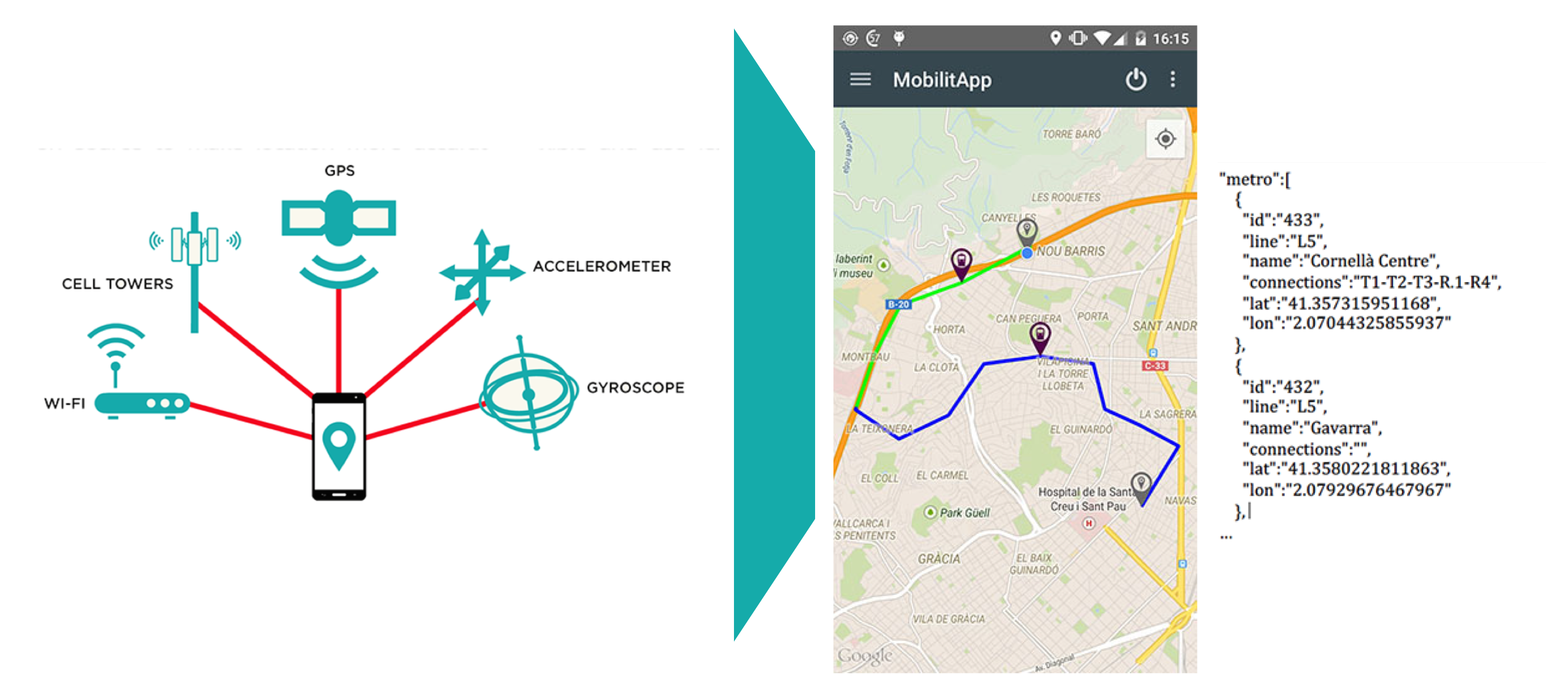}
\caption{Users' activities are collected considering different sources and sensors data. This information is processed to obtain citizens' mobility patterns.
\label{sources}}
\end{figure}

\subsection{Mobility patterns recognition}
MobilitApp is able to successfully classify between the following activity classes:

\begin{itemize}
 \item \emph{on foot}: Activity type returned if the citizen is either walking or running.
 \item \emph{bicycle}: Activity type returned if the citizen is on a bicycle.
 \item \emph{vehicle}: Activity type returned if the citizen is on a motor vehicle (e.g. car, motorbike, bus,...).
 \item \emph{still}: Activity type returned if the citizen is not moving.
 \item \emph{unknown}: Activity type returned if Activity Recognition API is not capable to estimate the actual activity.
\end{itemize}

A key challenge was to successfully classify different types of vehicles and distinguish between private and public transportation. We use a simplified geofencing technique to identify if the user is using public or private transportation. For example if we observe that a user has lost GPS contact while moving we might assume they have used the metro. Therefore we try to find the closest metro station from the first good GPS sample. The same technique could be applied to bus lines, while also considering other information such as timetables and average speed. However, with this approach we had to listen to GPS data with a very short interval, therefore consuming too much battery. Our next objective is relying more on acceleration data and less on GPS and positioning to identify users' activities.
\noindent
We are also considering the possibility to use alternative information to GPS, to estimate the user's position. We see promising the possibility to cross battery level and consumption information with signal strength and distance from known WIFI/cell access point~\cite{michalevsky2015powerspy}.

\subsection{Privacy conscious analytics}
We are aware that we are collecting sensitive user data constantly and this poses a security and privacy risk for the users participating in the program. We have therefore implemented a number of measures to safely store and analyse user data.  We follow the approach of Daniel J. Solove in ~\cite{solove2006taxonomy} to classify possible privacy violations in four main categories:
\begin{itemize}
  \item \emph{Collection}: Surveillance; Information probing; Interrogation.
  \item \emph{Processing}: Aggregation; Identification; Insecurity; Secondary use; Exclusion.							
  \item \emph{Dissemination}: Breach of confidentiality; Disclosure; Exposure; Increased accessibility; Appropriation; Distortion.
 \item \emph{Invasion}: Intrusion of someone's private life.												
\end{itemize}

To avoid exposing users to direct threats of \emph{collection} and \emph{processing} of private information, MobilitApp has the option not to supply any personal details to the platform. Users are not obliged to disclose their personal data. To avoid \emph{dissemination} and \emph{invasion}, user data collected by our mobile application is communicated encrypted to the server. We are aware that if an attacker would gain access to the MobiltApp platform they would be able to commit all possible privacy violation on user data. Said this we are continuing researching measures to reduce the users' privacy risk even in these circumstances.

\section{Conclusions and future work}
We are determined to continue developing MobilitApp and improve how we detect the user transportation mode and position. We will especially concentrate on how the phone accelerometer is used to detect the user's activity. More specifically the gravity component provided by the sensor can be estimated by the algorithm and then classified as one of the different mobility indicators that we can use. We want to emphasise that this new improvement will help us distinguish between different motorised transportation modes, which currently represent the main challenge for smartphone-based transportation mode detection.
Also, we are considering to implement methods for location discovery without the use of GPS, to reduce device battery consumtion. We find promising techniques using the transmitting and receiving power consumption together with distance between the user's device and the base station (access point/mobile cell).

\subsection*{Acknowledgments.} This work is supported by the Spanish Government through project INRISCO (INcident monitoRing In Smart COmmunities. QoS and Privacy, TEC2014-54335-C4-1-R). We are also grateful to Xavier Rossell\'o and Francesc Calvet from the Autoritat del Transport Metropolit\`a de Barcelona for their valuable feedback during different stages of the project.


\bibliographystyle{splncs03}
\bibliography{bibliography}

\end{document}